\documentclass[a4paper,11pt]{article}
\usepackage{jinstpub} 
\usepackage{color}
\usepackage{xcolor}

\title{\boldmath Direct Vertex Reconstruction of $\Lambda$ Baryons from Hits in CLAS12 using Graph Neural Networks} 

\author{Keegan Menkce, Matthew McEneaney, Anselm Vossen}
\affiliation{Lawrence University,\\
711 E Boldt Way, Appleton, WI 54911, USA}
\affiliation{Duke University,\\
2138 Campus Dr. Durham, NC 27708, USA}

\emailAdd{kmencke2@illinois.edu}

\abstract{
Machine learning techniques, including Graph Neural Networks (GNNs), have been used extensively for data analysis in high energy and nuclear physics. Here we report on the use of a GNN to reconstruct decay vertices of $\Lambda$ hyperons directly from hits in the tracking detector at the CLAS12 experiment at Jefferson Laboratory (JLab). We show that we can improve the  vertex reconstruction in simulation compared to the standard, track based, algorithm. We believe this warrants further study. The current study is limited by available training resources but points to an interesting possibility to forgo vertex reconstruction by track fitting in a complicated magnetic field for a more direct approach where the hit to vertex mapping is encoded in a neural network.
}


\begin{document}
\maketitle
\flushbottom

\section{Introduction}
\label{sec:intro}

The strong force governs interactions of the constituent quarks and gluons inside protons and neutrons. However, at the energy scales relevant for nucleon structure, the strong coupling constant makes a priori calculations of nuclear structure intractable~\cite{Collins_2011}. Therefore experimental measurements have to be used. A common type are scattering experiments where a point-like probe, e.g. an electron, is used to scatter off a proton. From the particles produced in the scattering reaction, one can learn about the dynamics of the strong force.
Among these particles $\Lambda$ hyperons play a special role, as their weak decay reveals their polarization~\cite{Global_lambda}. Thus, the spin structure of the $\Lambda$ is easily accessible from experiment. Here we consider the decay of the neutral $\Lambda$ into a negatively charged $\pi^-$ (hereafter referred to as $\pi$) and a proton. These decay products can be reconstructed in an experiment as they are relatively long-lived and create charged tracks that can be reconstructed in tracking detectors.
Since the $\Lambda$ decays weakly, its lifetime is relatively long~\cite{PDG} enabling it to travel typically on the order of centimeters, with a $c\tau\approx 8$~cm. Therefore the ability to find displaced decay vertices is essential when reconstructing $\Lambda$s from $\pi^-$ and proton candidate tracks.

The Continuous Electron Beam Accelerator Facility (CEBAF) at Jefferson Lab is able to achieve high enough energies to study nucleon structure. CEBAF sends high luminosity electron beam line to several experimental halls that are all fixed target experiments~\cite{doi:10.1146/annurev.nucl.51.101701.132327}. CLAS12 (CEBAF Large Acceptance Spectrometer at $12$~GeV) is located in Hall B of CEBAF. CLAS12 obtains excellent momentum and angular coverage. Measurements are made in the drift chambers, which determine particle trajectories~\cite{BURKERT2020163419}. The drift chambers are three regions (just before, inside, and just outside the magnetic torus field), with each region being broken up into six sections. Each region has wires in two super layers of six layers each. This forms a hexagonal pattern with field wires on the outside of the hexagon and sense wires at the center of each hexagon. In total there are 14 field wire layers and 6 sense wire layers ~\cite{MESTAYER2020163518}. The sense wires give the $x,y,z$ position of the particle, and the field wires give the direction the particle is moving in called $cx, cy, cz$. The three regions allow for the particle momentum to be reconstructed as the magnetic field is well known in this region.

For the vertex reconstruction we use the location of the hits in the drift chamber of the associated tracks. At least four hits are required.  Each hit in the chambers gives the position of the particle and the direction, allowing for the identification of particles' momentum~\cite{MESTAYER2020163518}.  
Here, we consider the process of production of $\Lambda$ in Semi-Inclusive Deep Inelastic Scattering (DIS), where an electron scatters off a target proton and the scattered electron and a $\Lambda$ are detected in the final state, $e+p\rightarrow e'+\Lambda+X$. Here $X$ describes the unobserved part of the final state. From the initial and scattered electron, the DIS variables $x$, $Q^2$ and $W$ can be reconstructed, where $x$ can be interpreted in the parton picture as giving the fraction of the struck parton carried of the parent proton, $Q^2$ is the 4-momentum squared transferred from the beam to the target and $W$ gives the mass of the final hadronic state. Using the momentum of the final state $\Lambda$, the variables $z_\Lambda$ and $x_F$ can be reconstructed, where $z_\Lambda$ is the momentum fraction outgoing quark the $\Lambda$ is carrying and $x_F$ is the so-called Feynman-x variable which is the ratio of the longitudinal $\Lambda$ momentum over the maximum longitudinal momentum, $p_\Lambda/p_{L_{max}}$. Positive $x_F$ values are selected to increase the probability of current fragmentation, i.e. that the detected $\Lambda$ is produced in the fragmentation of the outgoing quark. $\Lambda$ hyperons are unstable and thus have to be reconstructed from their decay. Here we consider the decay into $\pi$ and proton, $\Lambda \rightarrow \pi^- + p$. 
Precise vertex reconstruction to extract an improved $\Lambda$ signal by requiring a minimum flight distance and thus a lifetime of the parent particle. As the reaction system is boosted along the longitudinal direction of the beam, that is, the $z-$axis, only the vertex displacement along this axis is of interest here.

The current reconstruction relies on ``swimming'' the tracks back through the magnetic field to determine their most likely vertex. However, this is complicated by the fact that the magnetic field configuration close to the target is non-trivial as the fields of the magnets of the central tracker around the target and the forward tracker overlap. This complicates mostly the reconstruction of the $\pi$ vertex, as the $\pi$ produced in the $\Lambda$ decay is comparably slow. Given the fact that the magnetic field is well measured and simulated, we explore here a possible machine learning approach to improve the vertex reconstruction~\cite{thais2022graph}.

There have been some applications of Neural Networks (NNs) to secondary vertex reconstruction which improved on traditional analytical tracking methods~\cite{GOTO2023167836,Shlomi_2021}. Specifically, we are using a Graph Neural Network (GNN)~\cite{deep_learnin_and_graph_networks} which can be seen as a generalization of Convolutional Neural Network to operate on a graph like data structure. For our application, the benefit of GNNs is that they can handle varying size inputs that is difficult to be represented efficiently as a matrix but can be more naturally be represented as a graph which is the case for full physics event.

\begin{figure}[htbp]
\centering
\includegraphics[width=.4\textwidth]{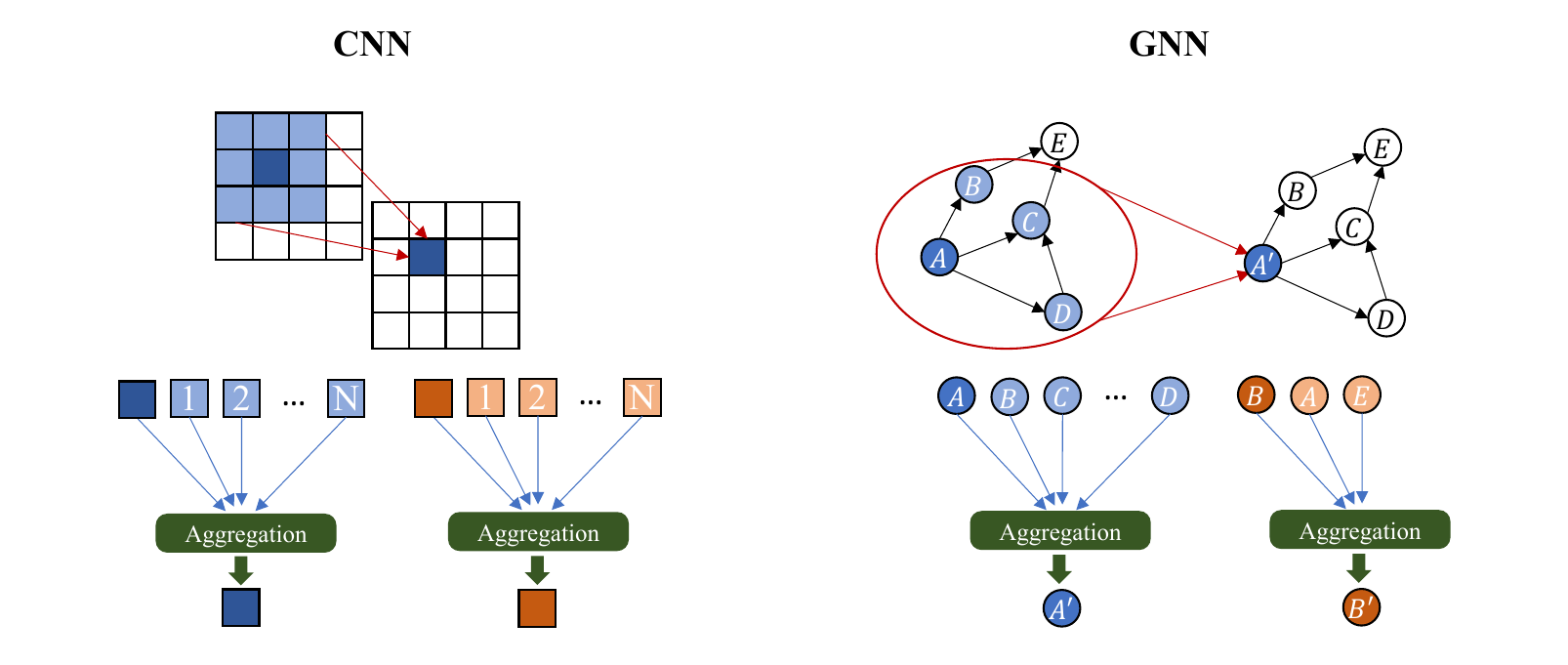} 
\caption{ Diagram showing the steps of the well know CNN compared with the GNN. In the CNN the neighbors are regular, while in the GNN the neighbors are defined by the graph. ~\cite{McEneaney_2023}}
\label{fig:CNNGNN} 
\end{figure} 

Here we describe our work to use a GNN to improve upon the resolution of $\pi$ longitudinal vertex in JLab's CLAS12 experiment. Using this we then use this vertex to improve the purity of the reconstructed $\Lambda$ sample. 

\section{Methods}  
\label{sec:methods}

We used a Graph Convolutional Network (GCN) implemented in pytorch ~\cite{NEURIPS2019_9015} with three convolutional layers, a dropout layer and a final linear layer. We used the standard Rectified Linear Unit (ReLU) as the activation function.

The data we used was obtained from simulations of $\Lambda$s produced in Semi-inclusive Deep-Inelastic scattering experiments with the CLAS12 experiment geometry. The input to the GNN was formed by forming a fully connected graph from the input of each track where the nodes are the hits of the decay proton and $\pi$s in the drift chamber that contain the six-dimensional hit information ($x,y,z,cx,cy,cz$). 
To give the algorithm a baseline, we also added the output of the existing vertex reconstruction that determined the vertex by looking for the closest approach between the track and the beamline as well as the reconstructed momentum. It was observed that regularization by the addition of Gaussian noise on the input vertex significantly improved the performance of the predictions.  The number of drift chamber hit inputs ranged from four to eight depending on the event. If an event had multiple proton and $\pi$ pairs, we choose the pair that was produced by a $\Lambda$ in the Monte Carlo simulations.

The network was trained to provide the  longitudinal components of the vertex of the track as output. For the supervised training the truth information from the CLAS12 Monte-Carlo simulations was used. We used simulated events with an reconstructed electron, proton and negative $\pi$ in the event with the electron being the trigger particle in the forward detector. In addition we  selected SIDIS events using constraints on the kinematics of the scattered lepton and the reconstructed hyperon. These are commonly used for other CLAS12 SIDIS analysis, with $Q^2>1GeV^2, W>2GeV, y<0.8, xF_{ppim}>0.0, mass_{ppim}<1.24GeV, z_{ppim}<1.0$. Here $ppim$ indicates the combined proton-$\pi^-$ system and thus the reconstructed $\Lambda$.  Finally, the composition of the training and test sample, was approximately $25\%$ and $50\%$ $\Lambda$, respectively.


Due to memory constraints on the available machines, we split the training set, which consists of about $8\times 10^4$ events in four sub-samples that we trained on separately for 200 epochs.
We then evaluated the results on a separate dataset of $20,000$ events. 
Compared to the available data the training set is limited and in the future, training with a larger set is desirable to fully cover the necessary phase space. 




\section{Results}
The goal of this work was the improvement of the reconstruction of $\Lambda$ decay vertex to use it as a selection criterion for $\Lambda$ hyperons and thus to be able to increase the signal purity. In the following, we compared our predictions and the predictions from the default algorithm for the longitudinal $\pi$ vertex  with the true vertex from simulations. 

\begin{figure}[htbp]
\centering
\includegraphics[width=.4\textwidth]{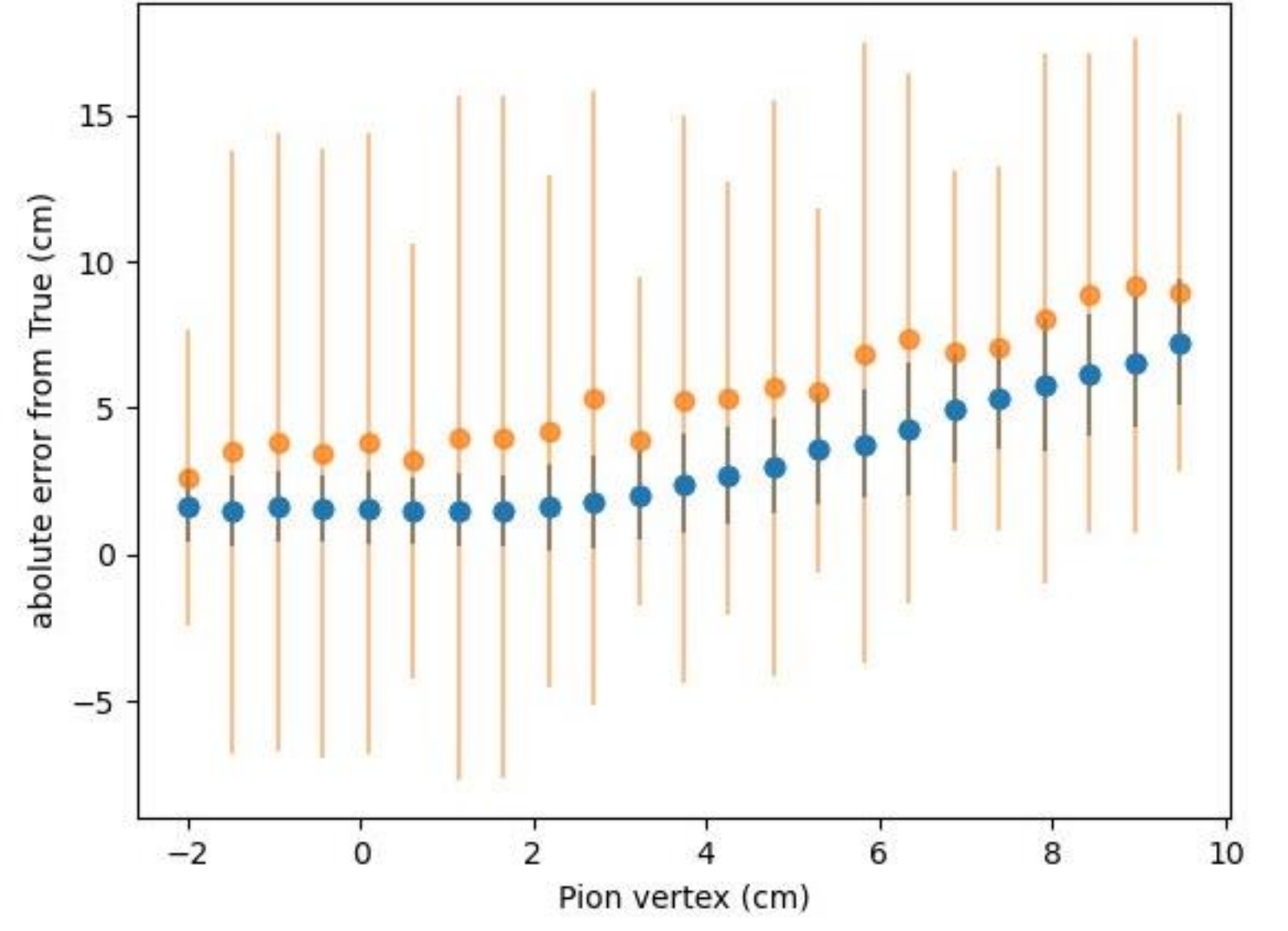}
\qquad
\includegraphics[width=.4\textwidth]{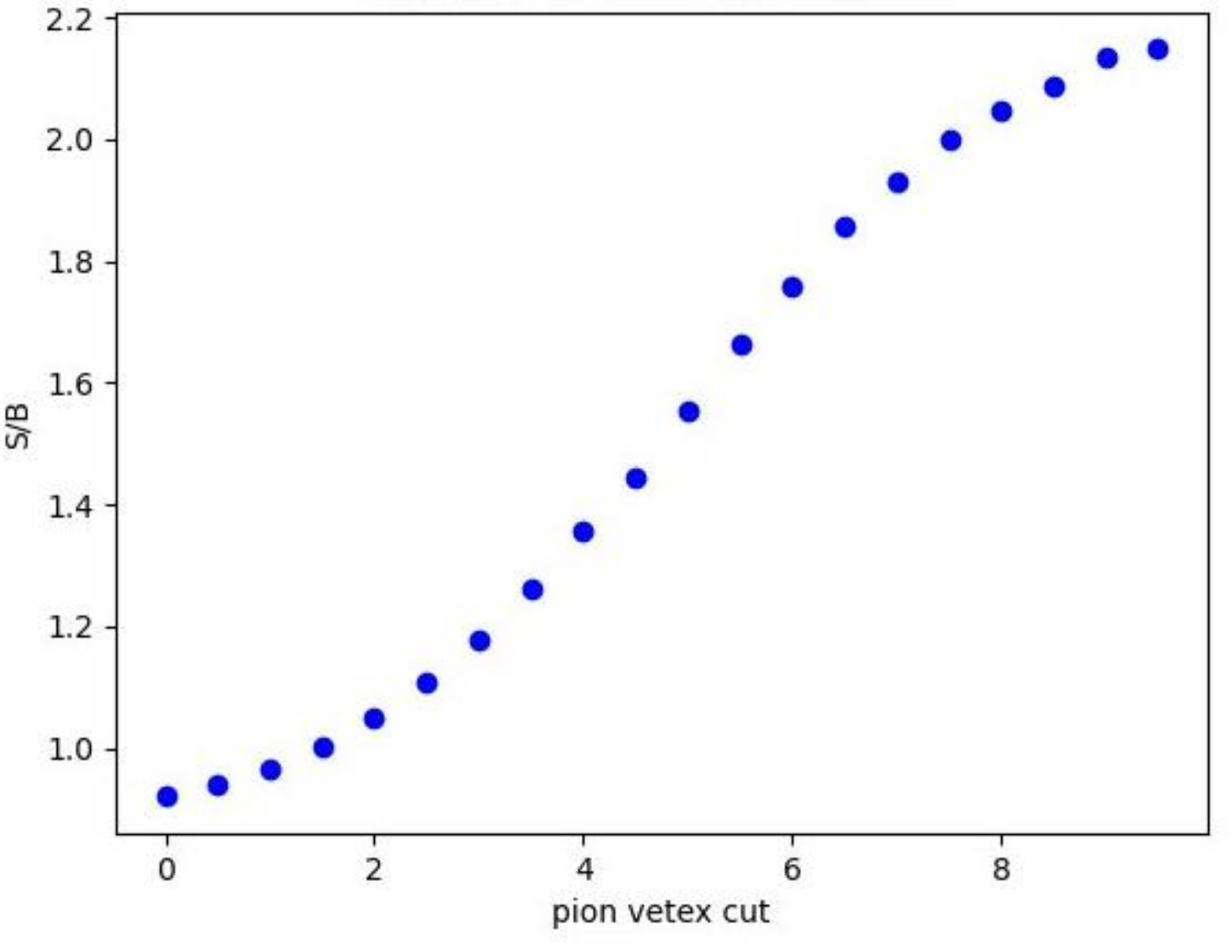}
\caption{Left. Mean absolute error based on true value vertex. Blue is the error for the GNN. Orange is the mean error for the standard algorithm. Error bars are the standard deviation of each. The x-axis is taken to be the true $\pi$ vertex. Right. Signal over background ratio as progressive cuts are made. Cuts were made based on GNN prediction of $\pi$ longitudinal vertex.  }
\label{fig:sb_GNN} 
\end{figure}

The right panel of Fig.~\ref{fig:sb_GNN} compares the absolute error for predictions using the GNN and  the standard algorithms versus the true $\pi$ vertex position. The absolute error is taken as $|prediction-truth|$. The errors at high true $\pi$ vertex's are especially important as those events are more likely to be $\Lambda$ events. The GNN prediction shows a lower absolute error across all true vertex values and has a lower standard deviation of this error. Although both error for both methods increase with the true vertex, the GNN prediction consistently remains lower.

The predicted $\pi$ vertex can be used to select $\Lambda$ hyperons based on requiring a minimum displacement of the vertex. This is shown in the right panel of Fig.~\ref{fig:sb_GNN}.  The panel shows dependence of the purity of the $\Lambda$ sample on the minimum required vertex.
As expected, the purity increases with tighter selection criteria.
Requiring a minimum vertex displacement of $0.0$cm brings above unity and a maximum purity is reached around $6.0$ cm where the purity reaches a value of just over $2.0$. 

\begin{figure}[htbp]
\centering
\includegraphics[width=.4\textwidth]{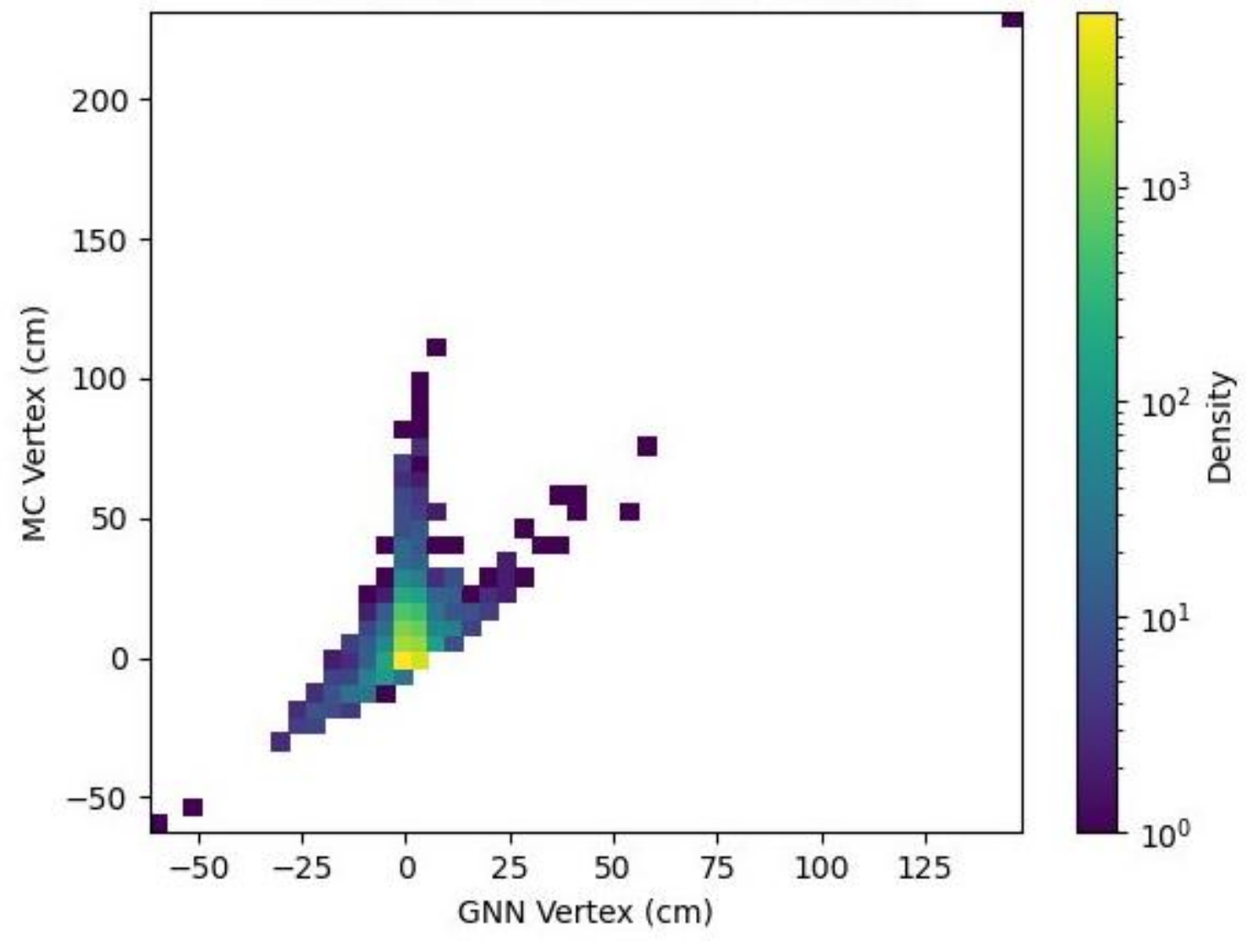}
\qquad
\includegraphics[width=.4\textwidth]{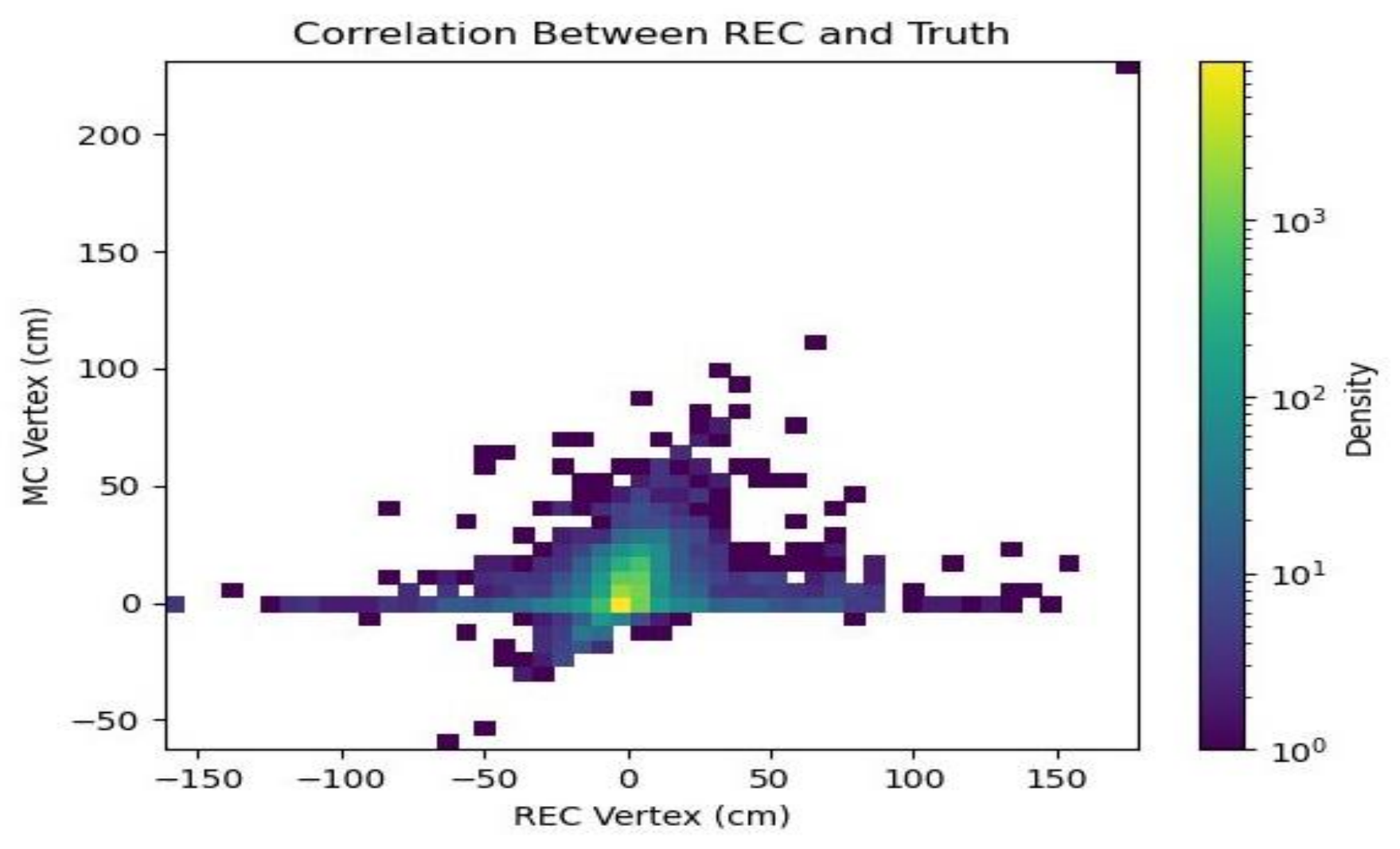}
\caption{Left: Correlation between GNN prediction and true vertex. Right: Correlation between vertex reconstructed with standard algorithm and true vertex.  }
\label{fig:scatter} 
\end{figure}

Figure~\ref{fig:scatter} illustrates the correlations between the outputs of the GNN and the standard reconstruction algorithm with the true vertex. The y-axis of both graphs is the true $\pi$ vertex. The x-axis of the left panel is the GNN prediction whereas it is the output of the standard algorithm in the right panel. As shown in Fig.~\ref{fig:scatter} for the GNN a clear correlation can be observed with the exception of small vertex displacements. This might reflect the true vertex resolution that can be achieved with the current tracking detectors or might be a remnant effect of the limited training data set.  In contrast, for the standard algorithm, the correlation between its output and the true MC vertex is much less pronounced.

\section{Discussion}
\label{sec:conclusion}

\begin{figure}[htbp]
\centering
\includegraphics[width=.4\textwidth]{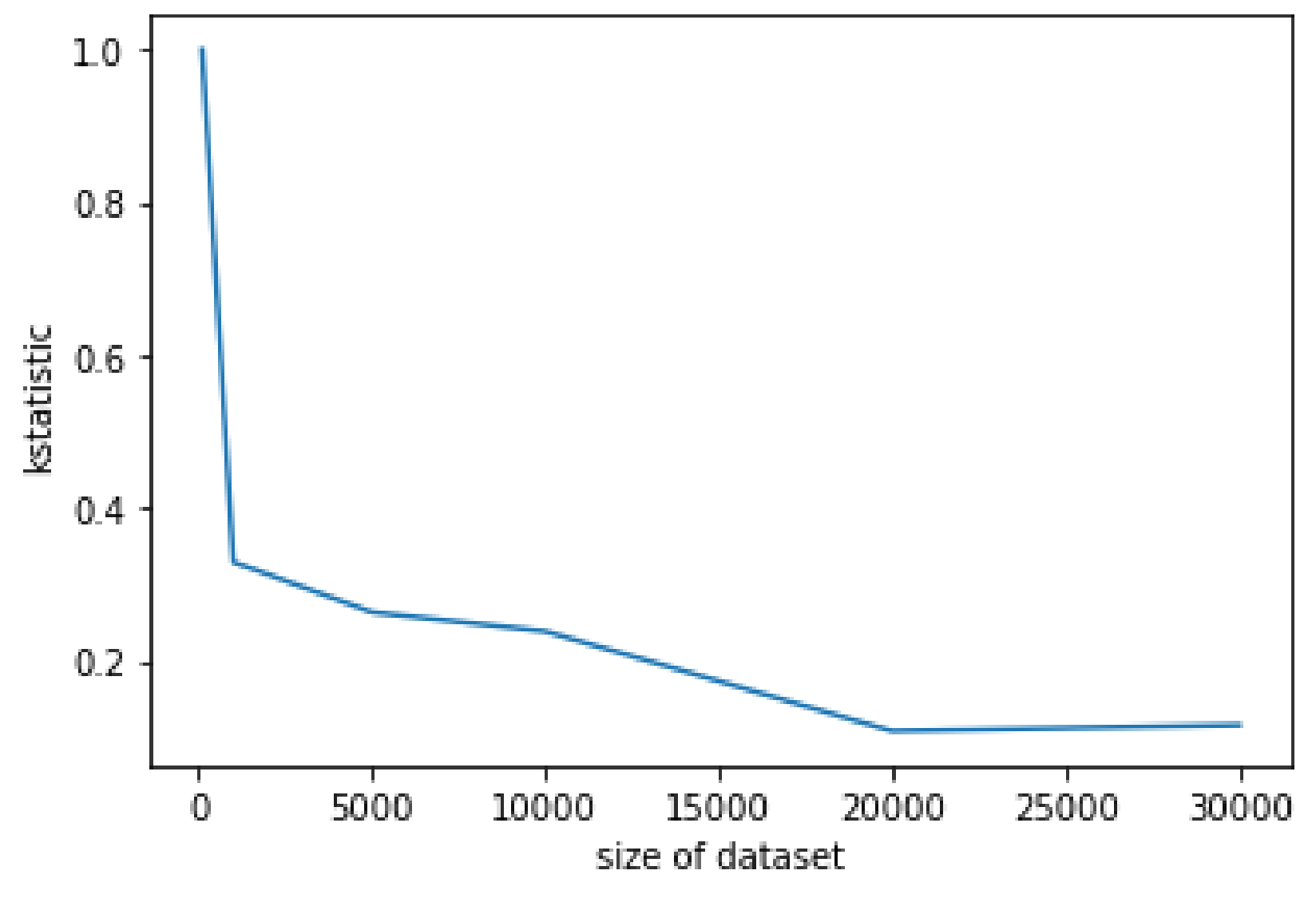}
\caption{ Improvement of Kolmogorov-Smirnov statistic as the size of the dataset increases. This shows that a bigger dataset does improve results. }
\label{fig:kstate} 
\end{figure}

Despite limited training and dataset size, our GNN showed an improvement in predicting the $\pi$ longitudinal vertex. It should be noted that CLAS12 has recently made an upgrade to the reconstruction algorithm that performs better, but we only tested our GNN against the older reconstruction algorithm. Performance could be enhanced by training for longer (more epochs), or utilizing a larger dataset as illustrated in Fig.~\ref{fig:kstate}. In the future, one could also investigate if, adding in information on the scattered electron can improve performance.

We demonstrated how this can be applied to the study of $\Lambda$. We showed promising results of GNN performance. Our GNN had less absolute error and a smaller standard deviation in the range $4-7$cm which is the region of interest to $\Lambda$ identification. Additionally, the GNN showed a stronger correlation with the true vertex compared to the reconstructed. The middle cluster is an indication of the precision limit that is possible from the hits. Finally, we show how cutting based on the GNN $\pi$ vertex can improve the signal to background ratio (s/b) ratio of $\Lambda$s.

To further improve these results, larger datasets to train on are essential. Fig.~\ref{fig:kstate} shows the dependence of the performance on the size of the training dataset. Another valuable  test would be to evaluate the results on a dataset with significantly less $\Lambda$ events to mimic experimental data. However, we opted against this as the statistics would be too low.  

In conclusion, we developed a GNN that predicts the z component of the $\pi$ vertex in JLabs's CLAS12 experiment. We choose to use a GNN architecture due to its ability to handle varying sizes of input data, like the data produced by the drift chambers in CLAS12. By using selections based on the $\pi$ vertex prediction, we improved the purity of our $\Lambda$ sample. These results are promising and we believe that further experimentation with larger and varied datasets is warranted to improve this method of increasing the resolution of $\Lambda$s with CLAS12.
\appendix

\acknowledgments

We acknowledge the outstanding efforts of the staff of the Accelerator, the Physics Divisions at Jefferson Lab, and the CLAS Collaboration in making this experiment possible

This material is supported by the U.S. Department of Energy, Office of
Science, Office of Nuclear Physics under Award Number DE-SC0024505 and also the National Science Foundation award PHY-2150118 (2022-2024).


\bibliographystyle{JHEP}
\bibliography{biblio.bib}

\providecommand{\href}[2]{#2}\begingroup\raggedright\begin{thebibliography}{10}

\bibitem{Collins_2011}
J.~Collins, \emph{Foundations of Perturbative QCD}, Cambridge Monographs on Particle Physics, Nuclear Physics and Cosmology, Cambridge University Press (2011).

\bibitem{Global_lambda}
{\scshape The STAR Collaboration} collaboration, \emph{{Global \ensuremath{\Lambda} hyperon polarization in nuclear collisions}}, \href{https://doi.org/https://doi.org/10.1038/nature23004}{\emph{Nature} {\bfseries 548} (2017) }.

\bibitem{PDG}
{\scshape Particle Data Group} collaboration, \emph{{Review of particle physics}}, \href{https://doi.org/10.1103/PhysRevD.110.030001}{\emph{Phys. Rev. D} {\bfseries 110} (2024) 030001}.

\bibitem{doi:10.1146/annurev.nucl.51.101701.132327}
C.W.~Leemann, D.R.~Douglas and G.A.~Krafft, \emph{The continuous electron beam accelerator facility: {CEBAF} at the {J}efferson {L}aboratory}, \href{https://doi.org/10.1146/annurev.nucl.51.101701.132327}{\emph{Annual Review of Nuclear and Particle Science} {\bfseries 51} (2001) 413} [\href{https://arxiv.org/abs/https://doi.org/10.1146/annurev.nucl.51.101701.132327}{{\ttfamily https://doi.org/10.1146/annurev.nucl.51.101701.132327}}].

\bibitem{BURKERT2020163419}
V.D.~Burkert and et~al., \emph{The {CLAS12} spectrometer at {J}efferson {L}aboratory}, \href{https://doi.org/https://doi.org/10.1016/j.nima.2020.163419}{\emph{Nuclear Instruments and Methods in Physics Research Section A: Accelerators, Spectrometers, Detectors and Associated Equipment} {\bfseries 959} (2020) 163419}.

\bibitem{MESTAYER2020163518}
M.~Mestayer, K.~Adhikari, R.~Bennett, S.~Bueltmann, T.~Chetry, S.~Christo et~al., \emph{The {CLAS12} drift chamber system}, \href{https://doi.org/https://doi.org/10.1016/j.nima.2020.163518}{\emph{Nuclear Instruments and Methods in Physics Research Section A: Accelerators, Spectrometers, Detectors and Associated Equipment} {\bfseries 959} (2020) 163518}.

\bibitem{thais2022graph}
S.~Thais, P.~Calafiura, G.~Chachamis, G.~DeZoort, J.~Duarte, S.~Ganguly et~al., \emph{Graph neural networks in particle physics: Implementations, innovations, and challenges},  2022.

\bibitem{GOTO2023167836}
K.~Goto, T.~Suehara, T.~Yoshioka, M.~Kurata, H.~Nagahara, Y.~Nakashima et~al., \emph{Development of a vertex finding algorithm using recurrent neural network}, \href{https://doi.org/https://doi.org/10.1016/j.nima.2022.167836}{\emph{Nuclear Instruments and Methods in Physics Research Section A: Accelerators, Spectrometers, Detectors and Associated Equipment} {\bfseries 1047} (2023) 167836}.

\bibitem{Shlomi_2021}
J.~Shlomi, S.~Ganguly, E.~Gross, K.~Cranmer, Y.~Lipman, H.~Serviansky et~al., \emph{Secondary vertex finding in jets with neural networks}, \href{https://doi.org/10.1140/epjc/s10052-021-09342-y}{\emph{The European Physical Journal C} {\bfseries 81} (2021) }.

\bibitem{deep_learnin_and_graph_networks}
P.W.~Battaglia, J.B.~Hamrick, V.~Bapst, A.~Sanchez-Gonzalez, V.~Zambaldi, M.~Malinowski et~al., \emph{Relational inductive biases, deep learning, and graph networks},  2018.

\bibitem{McEneaney_2023}
M.~McEneaney and A.~Vossen, \emph{Domain-adversarial graph neural networks for {$\Lambda$} hyperon identification with {CLAS12}}, \href{https://doi.org/10.1088/1748-0221/18/06/p06002}{\emph{Journal of Instrumentation} {\bfseries 18} (2023) P06002}.

\bibitem{NEURIPS2019_9015}
A.~Paszke, S.~Gross, F.~Massa, A.~Lerer, J.~Bradbury, G.~Chanan et~al., \emph{Pytorch: An imperative style, high-performance deep learning library},  in \emph{Advances in Neural Information Processing Systems 32}, pp.~8024--8035, Curran Associates, Inc. (2019), \href{http://papers.neurips.cc/paper/9015-pytorch-an-imperative-style-high-performance-deep-learning-library.pdf}{http://papers.neurips.cc/paper/9015-pytorch-an-imperative-style-high-performance-deep-learning-library.pdf}.

\end{thebibliography}\endgroup

\end{document}